%% file: workingpaper.tex
\documentclass[12pt,oneside]{article}
\usepackage[a4paper,left=2.5cm, right=2.5cm, top=2.5cm, bottom=2.5cm]{geometry}
\usepackage{authblk}
\usepackage{rotating}
\usepackage{setspace}
\usepackage[]{caption}
\usepackage{amsfonts}
\usepackage{amssymb}
\usepackage{amsmath}
\usepackage{bm}
\usepackage[T1]{fontenc}
\usepackage[utf8]{inputenc}
\usepackage{fouriernc}
\usepackage{graphicx}
\usepackage{placeins}
\usepackage{tabularx,booktabs}
\usepackage{longtable}
\usepackage{subfig}
\usepackage{rotating}
\usepackage{array,multirow}
\usepackage{tikz}
\usepackage{threeparttable}
\usepackage{natbib}
 \newcommand{\sym}[1]{$^{#1}$}
		
\makeatletter
\@fpsep\textheight
\makeatother	
\usepackage{csquotes}
\MakeOuterQuote{"}	
\title{Measuring Vacancies: Firm-level Evidence from Two Measures\thanks{We are indebted to Orley C. Ashenfelter, Paul Bingley, Per Krusell, Karsten Alb\ae{}k, John Hassler, Robert Shimer, Tobias Broer, Hannes Malmberg, Georg Marthin and Erik \"{O}berg for stimulating discussions and suggestions. We are very grateful to Charlotte Leolnar Reif, Anders Gabriel Pedersen, Klaus Henrik Langager and Charlotte Funk Christensen for providing us with data. All remaining errors are our own. Sievertsen acknowledges financial support from the Danish Council for Independent Research through grant 09-070295. Hansen acknowledges financial support from Institute for Evaluation of Labour Market and Education Policy and Handelsbanken's Research Foundations.}}

\author{Niels-Jakob Harbo Hansen \and Hans Henrik Sievertsen\thanks{Corresponding author. Univeristy of Bristol, VIVE, \& IZA. E-mail: \texttt{h.h.sievertsen@bristol.ac.uk}.} }

\date{This version: August 8, 2016.}
\begin{document}
\maketitle
\begin{minipage}{.9\linewidth}
\begin{abstract}
\onehalfspacing
Using firm-level survey- and register-data for both Sweden and Denmark we show systematic mis-measurement in both vacancy measures. While the register-based measure on the aggregate constitutes a quarter of the survey-based measure, the latter is not a super-set of the former. To obtain the full set of unique vacancies in these two databases, the number of survey vacancies should be multiplied by approximately 1.2. Importantly, this adjustment factor varies over time and across firm characteristics. Our findings have implications for both the search-matching literature and policy analysis based on vacancy measures: Observed changes in vacancies can be an outcome of changes in mis-measurement, and are not necessarily changes in the actual number of vacancies.
\end{abstract}  
\end{minipage}\vspace{24pt}
\clearpage

\section{Introduction}
\markboth{\MakeUppercase{Two Vacancy Measures}}{\MakeUppercase{Introduction}}
\label{sec:Intro}

%[Introduction]
In the empirical search-matching literature job openings are often measured using two alternative sources. The first is survey data on job vacancies compiled by statistical agencies. The second is register data on vacancies from databases maintained by Public Employment Services (PES). The former is often considered preferable, as it avoids the selection problems that are likely to be present in the latter \citep{Elsby2015}. However, since survey data for many countries, particularly in Europe, so far only is available for relatively short time periods the data from public job centers are used as proxy in many studies \citep{Berman1997, Carlsson2013, Hansen2004, Wall2002, Yashiv2000}.	

%[Research Question and Motivation]
An important, yet unanswered, question is how these two measures relate to each other on the micro level. This is important for a number of reasons. First, if we are interested in using the register based measure as proxy for the survey on a subset of the labor market, for which the survey is not available, we cannot simply rely on how well the aggregate times series track each other. Second, it is often assumed that the register based measure suffers from selection problems \cite{Elsby2015} write \enquote{a perennial concern is that such registered vacancies fail to be representative of all job openings}. These problems are likely to be smaller in the survey based measure. Yet it is also possible that the survey measure is not a super-set of the vacancies in the PES. That is, some vacancies are registered at the PES but not in the survey. Thus, by comparing the joint distribution on the micro-level of the two vacancy measures, we can obtain important information about the reliability of the data sources.

%[Method]
We investigate this question using firm level data for Sweden and Denmark. Specifically, we construct two new databases on the firm level. In the first database we match data for vacancies as measured in a survey conducted by Statistics Sweden with data for vacancies as counted in the database by the Swedish Public Employment Service. This database covers the period 2001-2012. In the second database we match data for vacancies as measured in a survey conducted by Statistics Denmark, with data for vacancies as counted in the database by the Danish Public Employment Service. This database covers the period 2010-2012. Using these databases we relate the two measures to each other across time and firm level characteristics.

%[Findings]
We find that on the the aggregate level, the number of vacancies in the PES accounts for roughly a quarter of all vacancies in the survey. This does not, however, imply that the vacancies in the survey constitute a superset of the PES vacancies. Specifically, we find that about 60\% of the vacancies in the PES are not included in the survey. To obtain the number of unique vacancies in the survey and PES, the survey needs to be adjusted by a factor of around 1.2. Furthermore, we find that this correction factor varies the over business cycle and with firm characteristics. This is important for both research on business cycles as it means that fluctuations in survey vacancies could be driven by changes in mis-measurement.

%Implications
These findings are also important in a methodological perspective. There appears to be consensus that surveys represent the best available  measures of vacancies \citep{Elsby2015,Shimer2005}.\footnote{\cite{Shimer2005} argues that the vacancy survey in the U.S. (JOLTS) provides \enquote{an ideal empirical definition} of job vacancies. \cite{Elsby2015} writes \enquote{Reliable, timely and comprehensive survey data have only recently been made available since the inception of the Job Openings and Labor Turnover Survey (JOLTS).}} However, our results suggest that reliance on survey measure is problematic. The extra 20\% of unique vacancies present in the PES provides a lower bound vacancies missing from the survey measure. This is one potential reason why a non-trivial share of hires are made with survey vacancies presented in the previous month. This phenomenon is documented by \cite{Davis2013} in the United States, and by Chapter 2 of this thesis for Sweden. 

%[Organization]
The paper proceeds as follows. In Section \ref{sec:paper3:data}, we describe our data sources and how the Danish and Swedish database are constructed. Section \ref{sec:firmlevel_analysis} documents the relationship between the two vacancy measures on the firm level. In Section \ref{sec:agg_implications} we compute the aggregate number of unique vacancies in the two databases and relate these to the aggregate number of vacancies in the survey. Section \ref{sec:paper3:conclusion} concludes.

\section{Data}
\markboth{\MakeUppercase{Two Vacancy Measures}}{\MakeUppercase{Data}}
\label{sec:paper3:data}

In this section we describe our data sources and how we construct our two new databases. We first describe our data on vacancies from surveys and then describe our data on vacancies measured by the Public Employment Services. Below we will refer to the first data source as \emph{survey data} and the second as \emph{register data}. 

\subsection{Data from vacancy surveys}

Our first data sources for vacancies are the Danish and Swedish \emph{Job Vacancy Survey}, respectively.

\subsubsection{Sweden}

The \textit{Swedish Job Vacancy Survey} is administered by Statistics Sweden and has been collected on a quarterly basis since 2001. Two vacancy concepts are available in this survey: (1) the number of available positions in each plant, which have been made \enquote{available for external job-seekers via the newspapers, internet or another mean of dissemination}, and (2) the number of these positions that the employer wishes to fill immediately.\footnote{In Swedish (1) is called \emph{Vakanser} and (2) is called \emph{Lediga jobb}.} This way, the former concept is a super-set of the latter. Below we rely on (1).

The \textit{Swedish Job Vacancy Survey} is collected at the plant level and all respondents are asked to report the number of vacancies in the middle of the reference month.\footnote{The respondents are asked to report the number of job openings on the Wednesday closest to the 15th of the reference month.} For the private sector the sampling is carried out on the plant level with approximately 16,700 work places sampled each period. For the public sector the sampling was also conducted on the plant level until the second quarter of 2006, when the sampling was changed to the organizational level and on this level 650 organizations are sampled each period. Units larger than 100 employees are asked to do the reporting for each month of the relevant quarter, whereas units with less than 100 employees only are asked to report in the reference month. Reporting takes place either via letter or online. Non-respondents are reminded via email, letter, or a phone call. Until 2004, reporting was voluntary and the share of non-reporting units roughly 30\%. In 2004 reporting became mandatory, and currently the share of non-reporting units is 11\% in the private sector and 2\% in the public sector.

\subsubsection{Denmark}

%[Concept]
The \textit{Danish Job Vacancy Survey} is administered by Statistics Denmark, and has been collected on quarterly basis since 2010 \citep{DST2016}. We have access to the data for the period 2010-12. The vacancy concept in this survey is a \enquote{position with salary, which is newly established, idle or about to be vacant, and that the employer is taking active steps to fill with a suitable candidate from outside the firm}. The plant is asked to report its number of vacancies on the second Wednesday of the given month.

%[Sampling]
The sampling is conducted on the plant level, with 7,000 plants surveyed each quarter. The plants are sampled from the \emph{Central Business Register}, which is a database covering all registered firms in Denmark. Data is reported on a quarterly basis but collected monthly with three equally sized groups reporting each month in the relevant quarter. Reporting firms are selected into one of these three groups randomly. Reporting is mandatory and plants that fail to report are first reminded and then informed that lack of reporting will lead to a police report. The share of non-reporting units is 5\%. For non-reporting units larger than 100 employees data is imputed.\footnote{Imputation is done for 2.5\% of all plants. This is accomplished  via prediction from a  regression model of vacancies on the number of employees within the same industry.}

That data is reported quarterly, but collected monthly, which constitutes a challenge. Indeed, for a given observation we know only the quarter for which the data was collected, while the exact month is unknown to us. We thus assume that the observed number of observations is representative for all months in the particular quarter. Specifically, we compare the observed number of survey vacancies to the \emph{average} number of PES vacancies in the relevant quarter.

\subsection{Data from Public Employment Service}

Our second data source for vacancies is the Danish and Swedish register data from the Public Employment Service (PES). 

\subsubsection{Sweden}

For Sweden, we have access to all vacancies registered at the \textit{Swedish Public Employment Service}. This database covers the all job vacancies made at the agency during the period 2001-2012. The database contains a row for each posting made at the agency with information on the start and end date of the posting along with information on the number of workers the firm is searching for and information on job and firm type. In principle, the database contains both a firm and plant identifier. However, for the majority of the observations the plant identifiers are missing, so the database in practice only contains useful information on the firm level. The database is known to contain a number of duplicates. To remove these, we drop vacancies that are identical to another vacancy in firm identifier, start-date, and job-characteristics.\footnote{This operation removes 8\% of the rows in the database. One row can contain more than one vacancy.}  Furthermore, to allow for comparison to the survey, we restrict our sample to the PES vacancies that are active on the reference day of the survey. 

\subsubsection{Denmark}

For Denmark, we use all vacancies announced at the \textit{Danish Public Employment Service} throughout the period 2010-2012. Each row in the database contains a firm identifier, the number of positions available, on the announcement day of the vacancy as well as the date on the vacancy was withdrawn from the database. As in the Swedish case we limit our sample to vacancies active on the reference day of the survey.

\subsection{Background variables}

For both Denmark and Sweden, we have access to background variables at the firm level. 

For Sweden we have access to information on industry, firm size, value added, and turnover. The vacancy survey provides industry information on the NACE level, and using this data we classify the firms into eight industry groups.\footnote{The Statistical classification of economic activities in the European Community is abbreviated as NACE.} We further have access to a subset of the variables from the company section of the \emph{Longitudinal Integration Database for Health Insurance and Labour Market Studies} (LISA). This is a register database maintained by Statistics Sweden, that provides information on firm size, value added and turnover for the registered firms at an annual frequency. 

For Denmark we have access to industry information, firm size and average wage level on the firm level. All these data derive from \emph{The Integrated Database for Employment Research} (IDA), which is developed and maintained by Statistics Denmark. The database contains the set of firms where at least one person is registered as being employed during November each year.

\subsection{Data selection}

The survey data is collected on the plant level, while the register data only has useful identifiers on the firm level. To address this issue, we  restrict our sample to firms with only one plant. For Denmark we obtain the number of plants per firm from the IDA database. For Sweden the number is obtained from the IFAU database, which contains information on all employment spells on the plant and firm level. This data is based on Swedish tax records, and allows us to compare the correct number of survey and register vacancies to each other, but it clearly comes at the cost of restricting attention to a non-random subset of the data. In Table \ref{tab:sampleselect}, we show that 43 and 35\% of all observations for Sweden and Denmark, respectively, are lost through this operation.

Analyses that exploit firm background variables further reduces the size of the sample. While the reduction is negligible in DK, the sample is reduced by 14\% in the Swedish case.

In Table \ref{tab:representiveness} we assess the representativeness of our sample \emph{vis-a-vis} the distributions of firms in the wider economy. Table \ref{tab:representiveness} shows that \emph{Farming, fishery, and mining} is over-sampled in both countries, while \emph{manufacturing} is under-sampled. Besides that, the distribution across industries roughly mimics that in the economy. Across size, our sample has a an over-representation of larger firms, while firms with high value added (for Sweden)/wage (for Denmark) per worker are over-sampled. This may seem at odds with the data restriction to firms with only one plant, but larger firms are likely to have been over-sampled initially in the vacancy surveys.  

Finally, for a fraction  of observations in the PES data there missing firm identifiers. For Denmark this share is low (1.5\%) over the entire sample period, while for Sweden it falls from around 30\% in 2001 to 1\% in 2010 and remains constant thereafter.\footnote{The exact shares are 32.6\% in 2001, 26.2\% in 2002, 17.8\% in 2003, 13.25\% in 2004, 10.5\% in 2005, 8.4\% in 2006, 6.6\% in 2007, 3.4\% in 2008, 2.8\% in 2009, 1.1\% in 2010, 1.2\% in 2011 and 1.1\% in 2012. } However, our conclusions are robust to restricting the Swedish sample to the period after 2009.\footnote{To check for the impact of this issue we have redone our main table for Sweden (Table \ref{tab:regs}) with a sample restricted to after 2009 in the Table \ref{tab:regs_before_2009}.}

% TABLE 1 SAMPLE SELECTION
\begin{table}[!ht]
\begin{center}
			\caption{Sample selection}
			\label{tab:sampleselect}
		\begin{minipage}{.7\linewidth}
\begin{flushright}
			\begin{tabularx}{1\linewidth}{Xccc}
		&\multicolumn{3}{c}{SWE}\\
		&N&Frac&Vac\\
		\toprule
	\input{tab_selection_SWE}
		\bottomrule
		\end{tabularx}
\end{flushright}
	\end{minipage}\hspace{-0.12cm}
		\begin{minipage}{.28\linewidth}
		\begin{tabularx}{1\linewidth}{ccc}
		\multicolumn{3}{c}{DEN}\\
		N&Frac&Vac\\
		\toprule
		\input{tab_selection_DEN}
		\bottomrule
		\end{tabularx}
	\end{minipage}
\end{center}
\end{table}

\subsection{Descriptive statistics}
\begin{figure}[!ht]
	\subfloat[Sweden]{\includegraphics[width=0.5\linewidth]{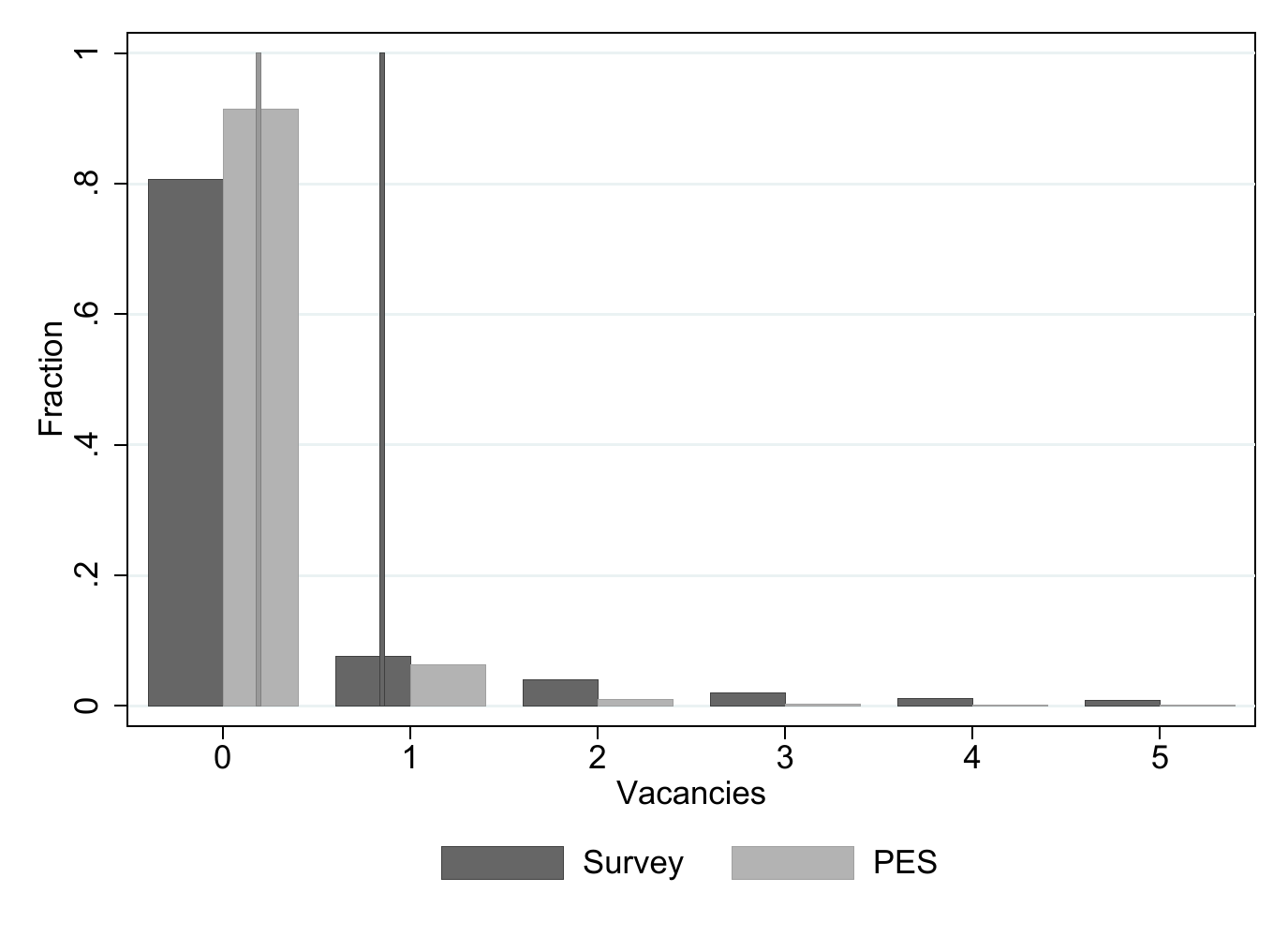}}
	\subfloat[Denmark]{\includegraphics[width=0.5\linewidth]{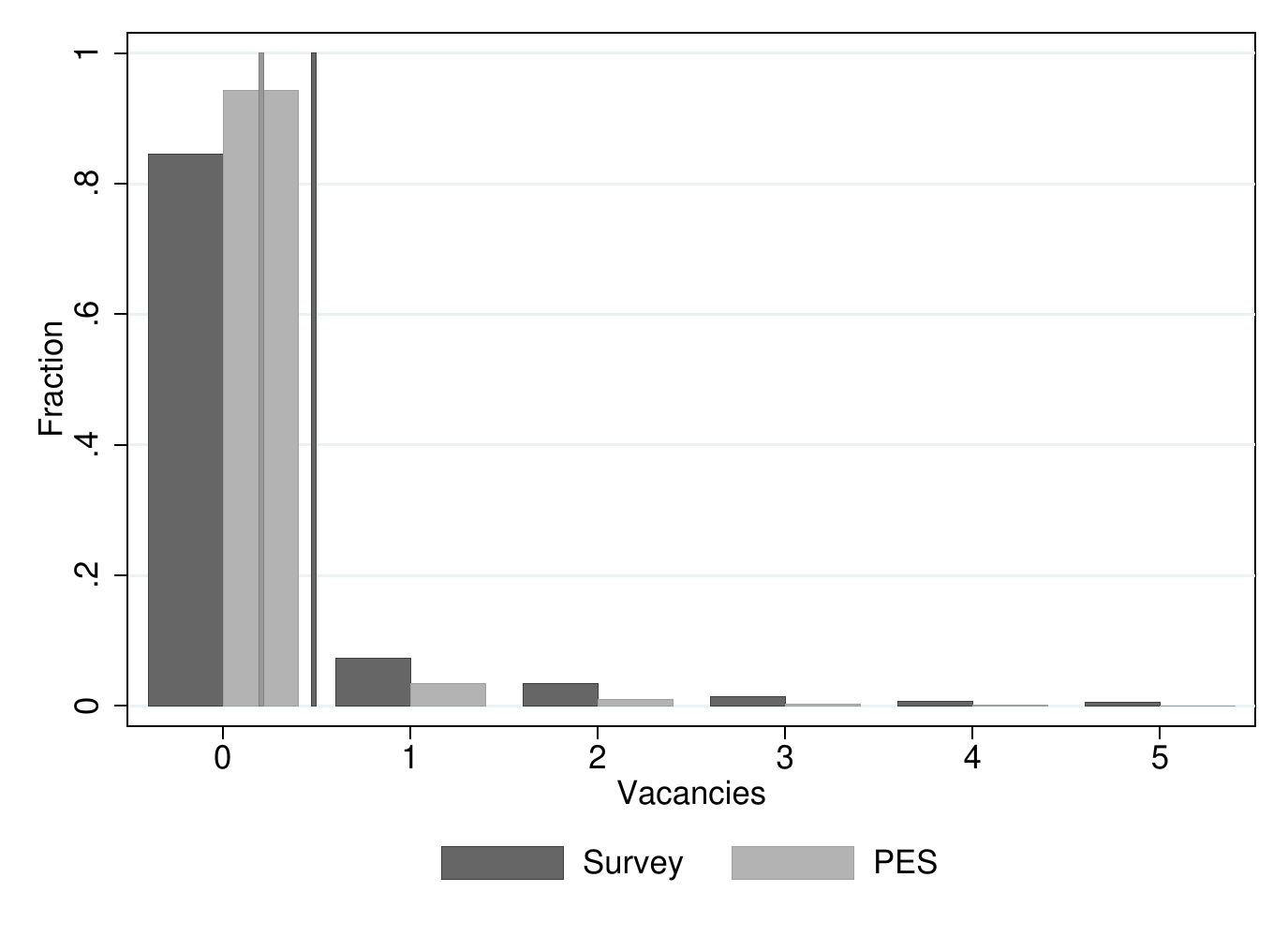}}\\
	\caption{Vacancy distribution in the Survey and the Public Employment Services data (PES)}
	\label{fig:hist_vac}
\end{figure}

Figure \ref{fig:hist_vac} shows the distribution of vacancies for both measures and countries. We note three similarities between the two countries. One, is that  more than 80\% of the observations have zero vacancies. A second is that the fraction of observations with zero vacancies is somewhat higher in the PES than in the survey. The third similarity is that the fraction of observations is monotonically decreasing in the number of vacancies. There are, however, also a dissimilarity between the countries: the average of survey vacancies is 75\% higher in the Swedish data. 

% Figure 1 Histograms

Table \ref{tab:descriptives} presents means and standard deviations for the background variables. The distribution of firms across both industries and sizes differs considerably across the two samples.

% TABLE 2 DESCRIPTIVES
\begin{table}[!ht]
		\begin{center}
			\caption{Descriptive Statistics}
			\label{tab:descriptives}
		\begin{minipage}{.75\linewidth}
	\begin{flushright}
			\begin{tabularx}{1\linewidth}{cXccc}
			\toprule
		&&\multicolumn{2}{c}{SWE}\\
		&&Mean&SD\\
		\toprule
		\multicolumn{4}{l}{\emph{Industry}}\\
	\input{tab_descriptives_SWE}
		\bottomrule
		\end{tabularx}
	\end{flushright}
	\end{minipage}\hspace{-0.12cm}
		\begin{minipage}{.2\linewidth}
		\begin{tabularx}{1\linewidth}{cc}
		\toprule
		\multicolumn{2}{c}{DEN}\\
		Mean&SD\\
		\toprule
		\input{tab_descriptives_DEN}
		\bottomrule
		\end{tabularx}
	\end{minipage}
	\begin{minipage}{.95\linewidth}
	\footnotesize{Notes: Value-added per worker and gross salary per worker is adjusted to the 2012 price level using the harmonized consumer price index for both countries and converted to EUR using the 2012 exchange rates.}	
	\end{minipage}
	\end{center}
\end{table}

\section{Patterns at the firm level}
\markboth{\MakeUppercase{Two Vacancy Measures}}{\MakeUppercase{Patterns at the firm level}}
\label{sec:firmlevel_analysis}

\begin{figure}[!ht]
	\subfloat[Sweden]{\includegraphics[width=0.5\linewidth]{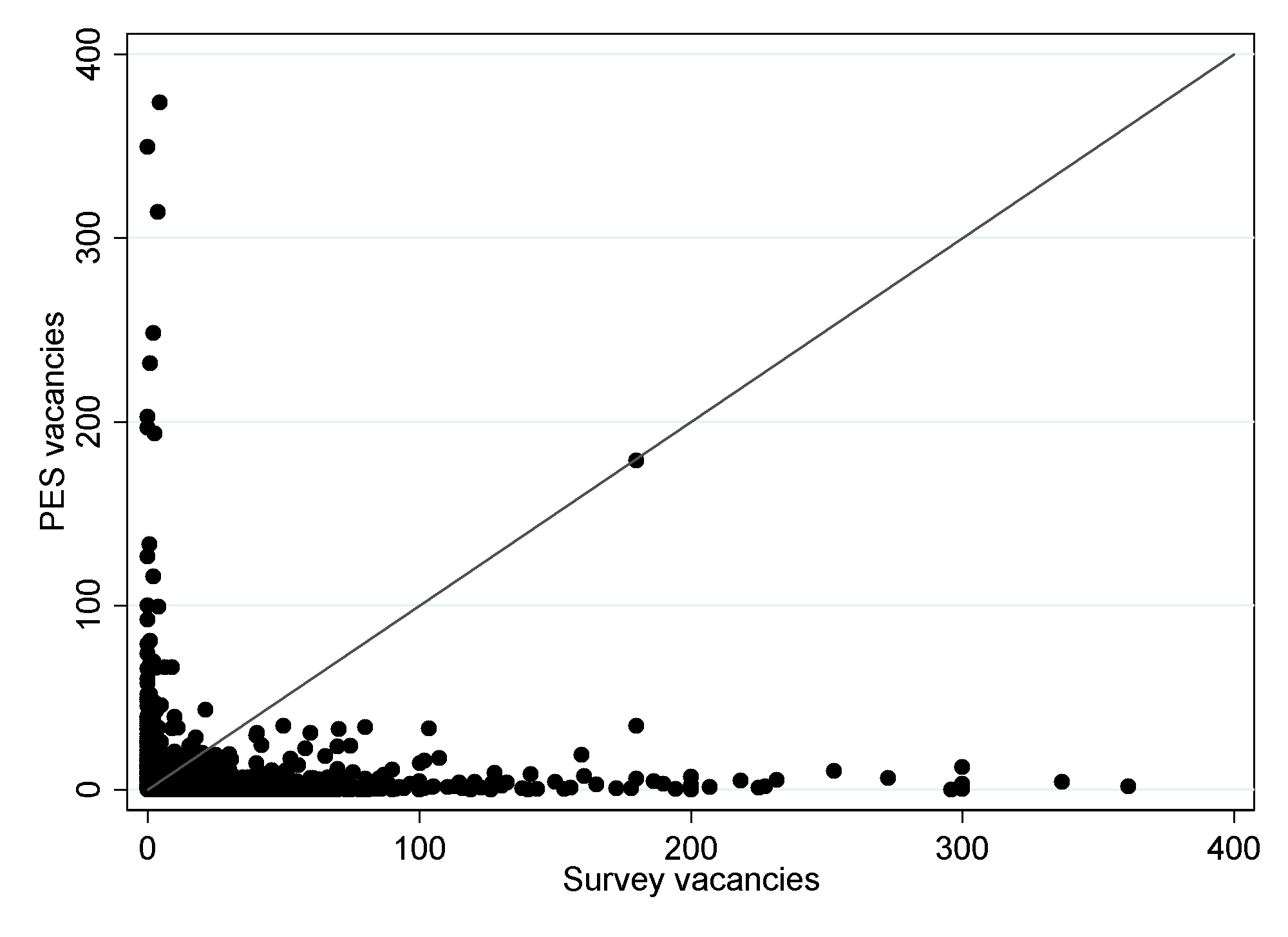}}
	\subfloat[Denmark]{\includegraphics[width=0.5\linewidth]{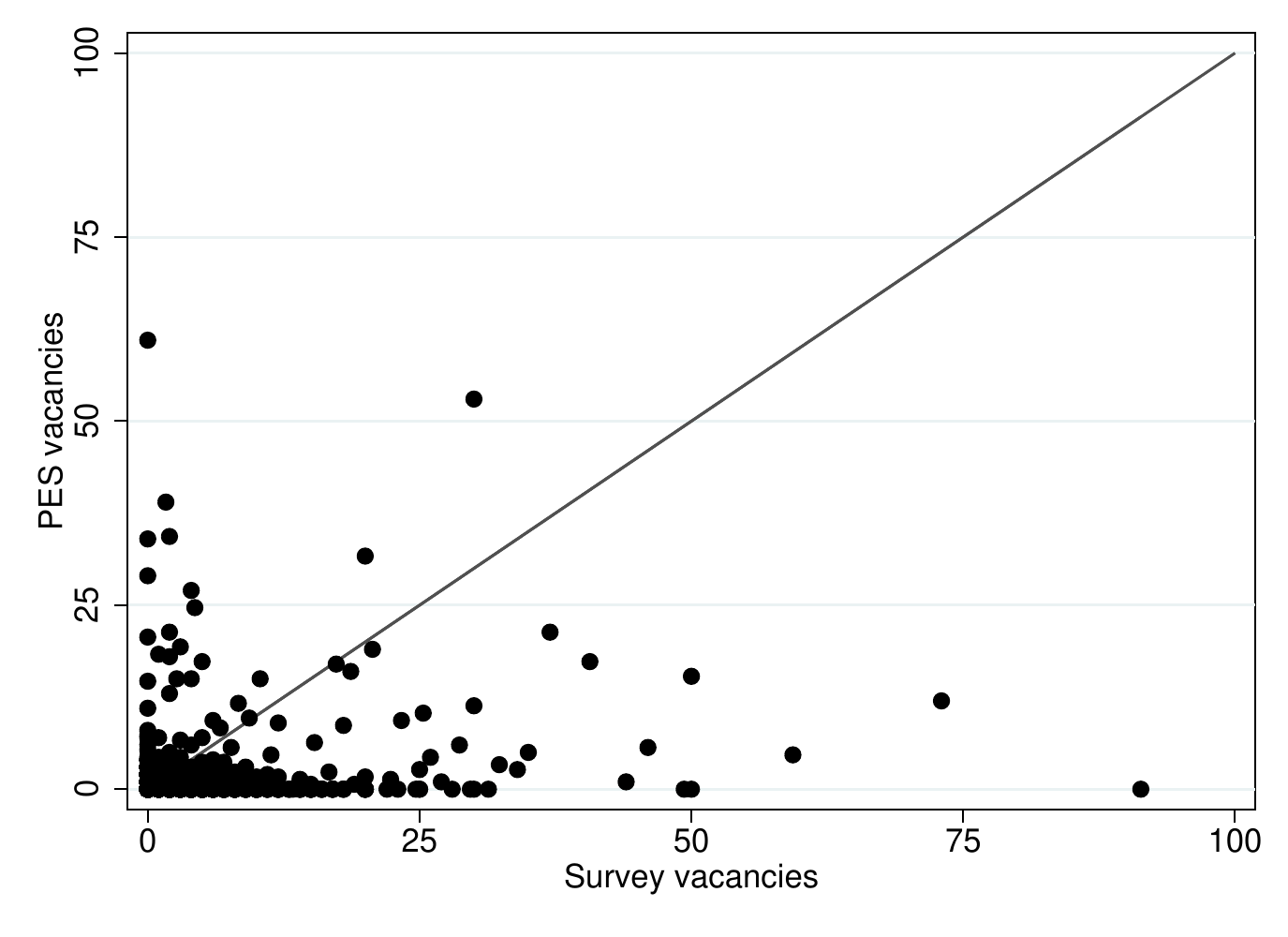}}\\
	\caption{PES vacancies plotted against survey vacancies for Sweden and Denmark. Each dot represents the average across three firms.}
	\label{fig:scatterplot}
\end{figure}

Figure \ref{fig:scatterplot} shows scatter plots of vacancies as measured in the survey and in the PES on the firm level. One dot represents the number of vacancies in the survey. For the PES, a dot is an average over three firms for one point in time.\footnote{The averaging is carried for anonymity reasons, but does not affect the patterns in the figures.} Two observations are clear from this figure. First, the dots are not scattered around the 45-degree line, which would be the case if the two measures represented the same vacancy concept. Second, there are a substantial number of observations above the 45-degree line. This is surprising as one would expect the survey data to be a super-set of the data from the PES, following the definitions in Section \ref{sec:paper3:data}.

% Figure 2 scatter plots

The interpretation of the scatter plot could be clouded by the substantial number of observations around zero. However, regression results confirm that the slope coefficient is considerably smaller than one.\footnote{These regressions are available upon request.}

% Figure 2 heatmaps
\begin{figure}[!ht]
	\subfloat[By survey columns, Sweden]{\includegraphics[width=0.5\linewidth]{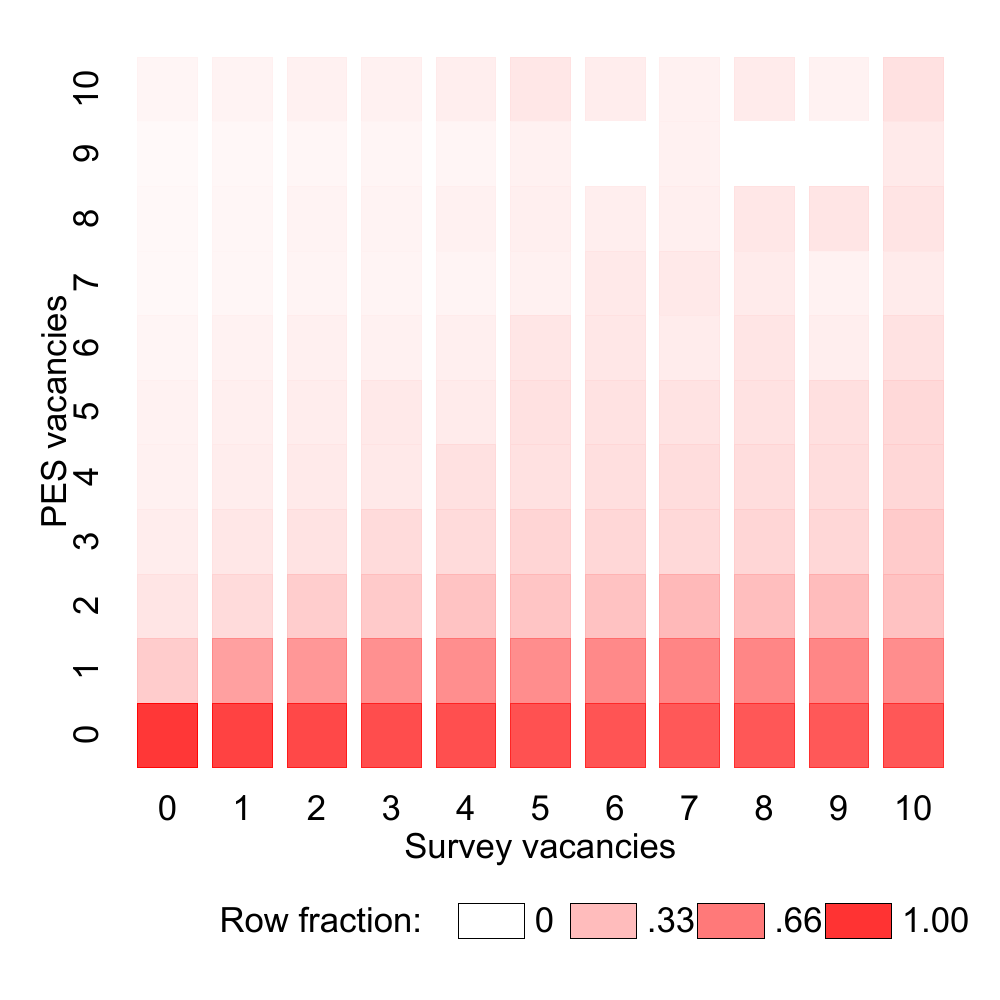}}
	\subfloat[By survey columns, Denmark]{\includegraphics[width=0.5\linewidth]{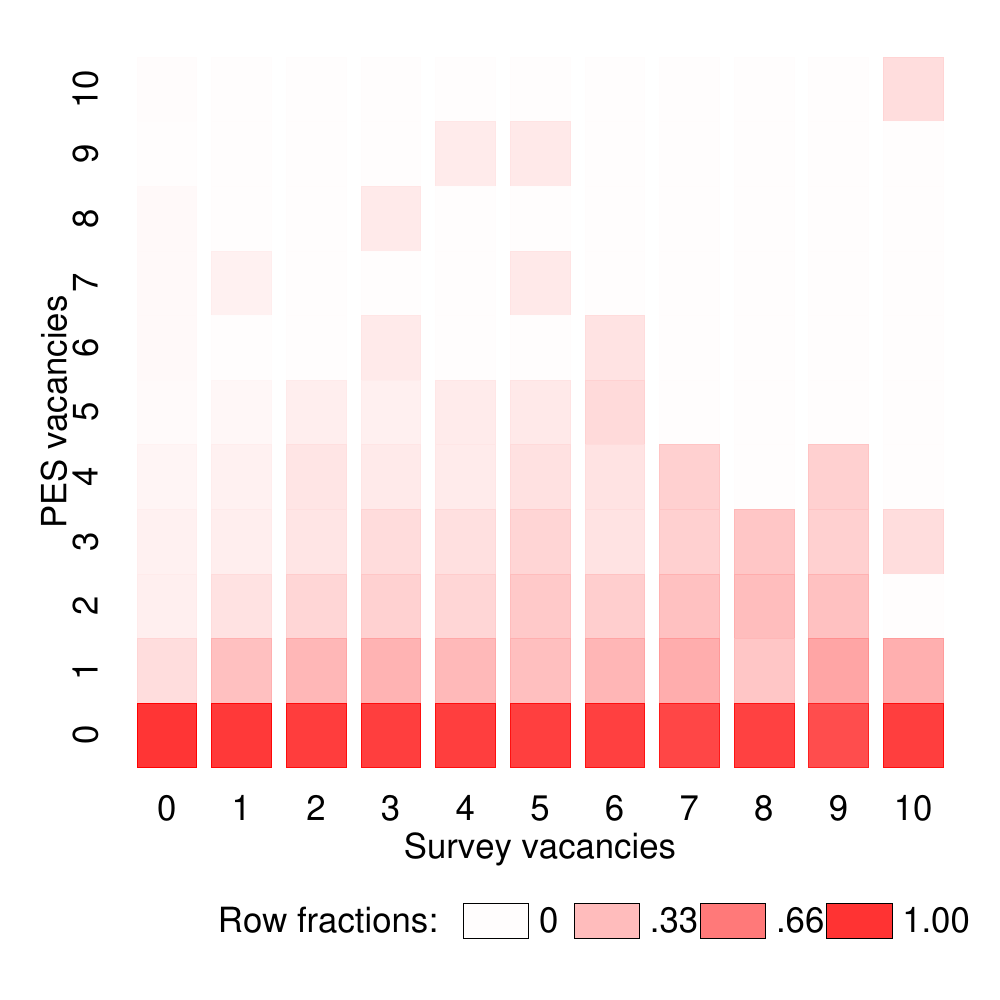}}\\
	\subfloat[By PES rows, Sweden]{\includegraphics[width=0.5\linewidth]{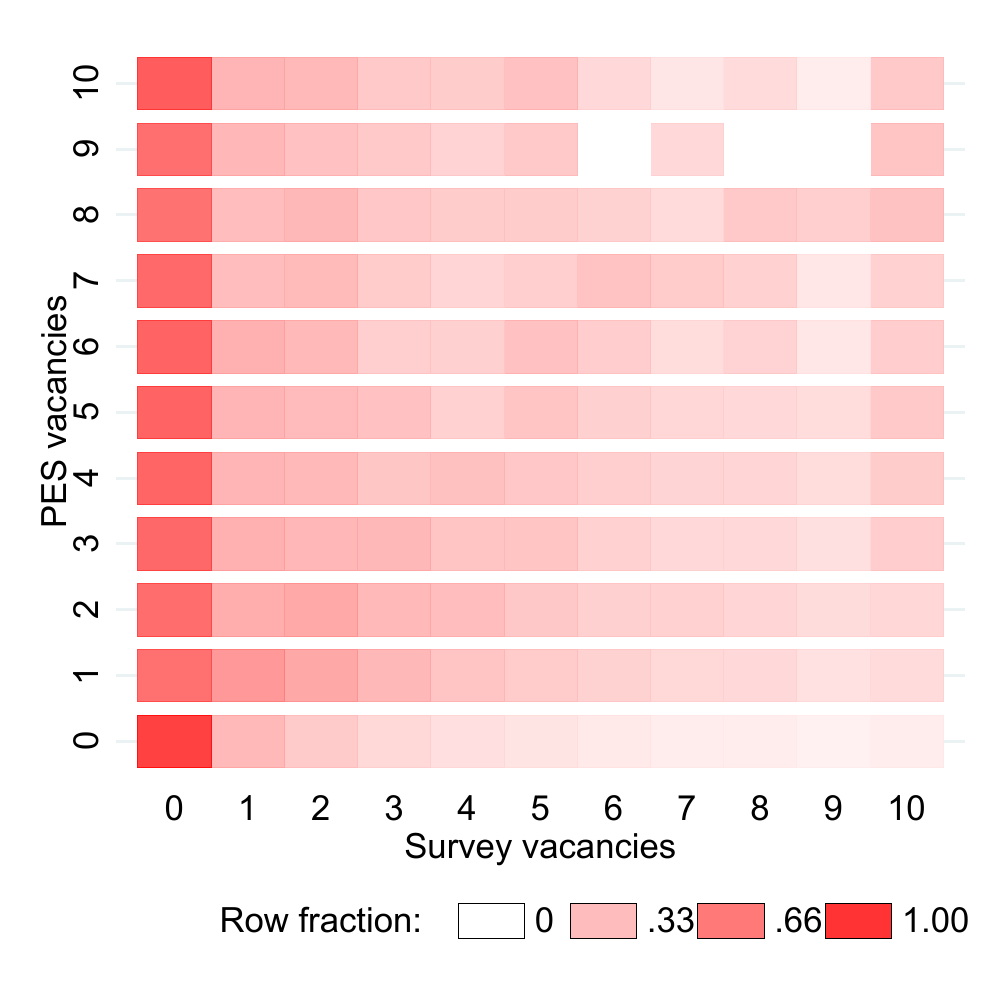}}
	\subfloat[By PES rows, Denmark]{\includegraphics[width=0.5\linewidth]{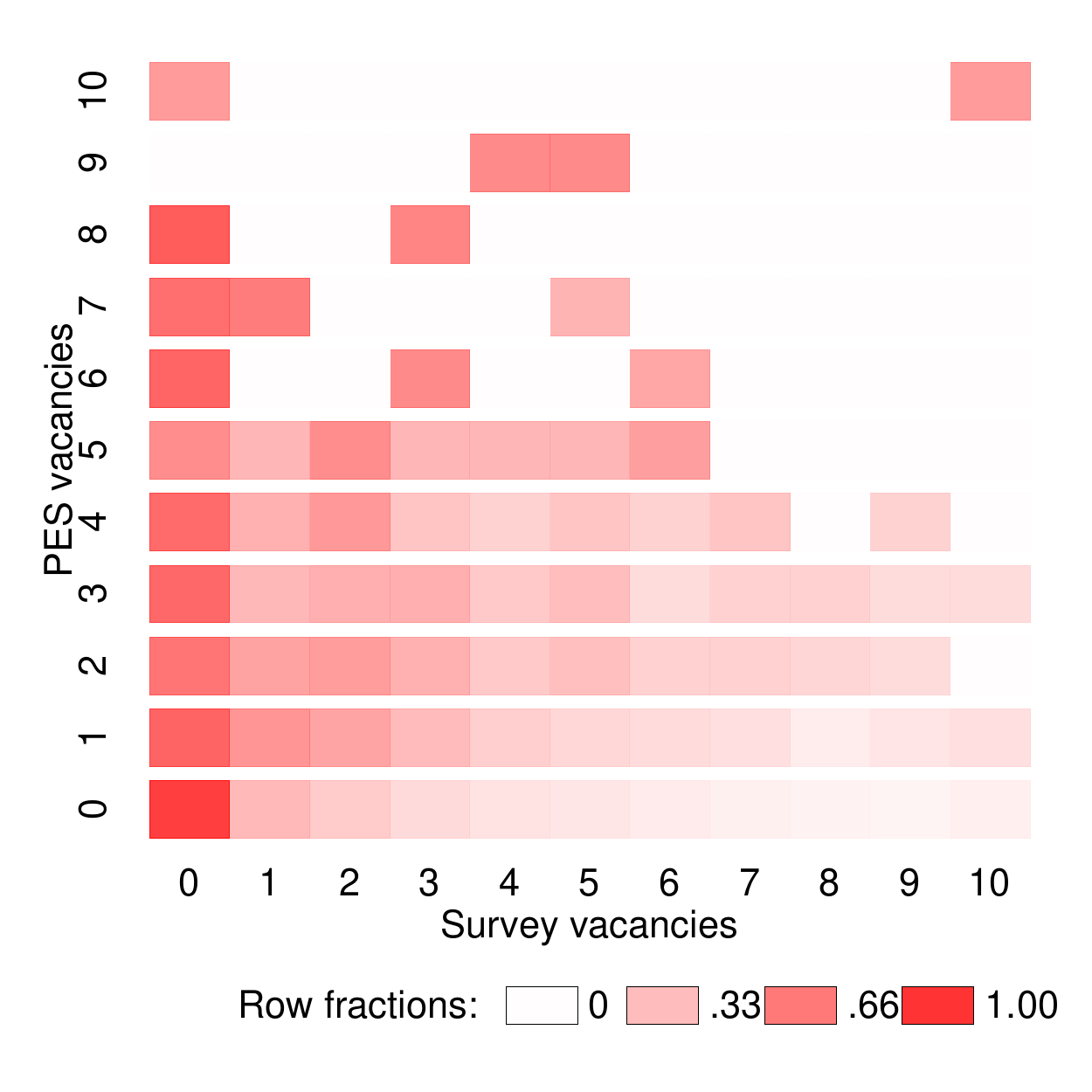}}\\
	\caption{Distributions of PES vacancies for given levels of survey vacancies and vice versa.  }
	\label{fig:heatmaps}
\end{figure}

In Figure \ref{fig:heatmaps} we look more closely at the joint distribution of the PES and survey vacancies. Panels (a) and (b) show the concentration of PES vacancies for a given number of survey vacancies. From these graphs we see that regardless of the number of observations registered in the survey, most of the density mass is concentrated at zero PES vacancies. However, the variation in PES vacancies increases in the number of survey vacancies. Panels (c) and (d) show the same graphs, but with the concentration of survey vacancies for a given number of PES vacancies. A similar pattern is observed here. The pattern is clearer for Sweden, which is likely due to the larger sample size. Table \ref{tab:rowcols} also presents the joint distribution. We note that around 80\% of the observations have both zero PES and survey vacancies, 18\% (15\%) of the observations have more survey than PES vacancies in Sweden (Denmark), and 4\% (2\%) have more PES than survey vacancies. 

% TABLE 4 Row and col elements
\begin{table}[!ht]
\begin{center}
			\caption{Row and Column Elements}
			\label{tab:rowcols}
\resizebox{1\linewidth}{!}{
				\begin{minipage}{.56\linewidth}
		\begin{center}
		Sweden\vspace{12pt}\\
		\begin{tabularx}{1\linewidth}{cc|cccccc}
		\multicolumn{2}{c}{}&\multicolumn{4}{c}{Survey}\\
		  && 0&1&2&3&4+\\
  \toprule
\parbox[t]{2mm}{\multirow{4}{*}{\rotatebox[origin=c]{90}{PES}}}
		\input{tab_vac_survey_PES_cross_SWE}
		\end{tabularx}
		\end{center}
	\end{minipage}
		\begin{minipage}{.56\linewidth}
		\begin{center}
		Denmark\vspace{12pt}\\
		\begin{tabularx}{1\linewidth}{cc|cccccc}
				\multicolumn{2}{c}{}&\multicolumn{5}{c}{Survey}\\
		  && 0&1&2&3&4+\\
  \toprule
%\parbox[t]{2mm}{\multirow{5}{*}{\rotatebox[origin=c]{90}{PES}}}
		\input{tab_vac_PES_SURV_DEN}
		\end{tabularx}
		\end{center}
	\end{minipage}
}
\end{center}
\end{table}

In Table \ref{tab:fracshares} we show the ratio of PES to survey vacancies (column $P/S$). Columns $PS$ and $SS$ in the table show the relevant shares by subgroups of total PES and survey vacancies, respectively. Four observations can be made from this table. First, the aggregate share of PES to survey vacancies is around one quarter in both countries. Second, while \emph{Farming, fishery and mining} is the industry with the highest fraction in Sweden, \emph{Public and personal services} is the industry with the highest fraction in Denmark. Note however, that \emph{Farming, fishery and mining} only accounts for a small share of both survey and PES vacancies. Third, for Sweden the fraction of PES to survey vacancies is decreasing in firm size, while no clear pattern is visible in the Danish data. Fourth, for Sweden we observe that the share is decreasing in value added per worker, while for Denmark the share is decreasing in salary per worker.

% TABLE 5  Fractions
\begin{table}[ht!]
\begin{center}
			\caption{Fractions}
			\label{tab:fracshares}
		\begin{minipage}{.73\linewidth}
			\begin{tabularx}{1\linewidth}{cXcccc}
		&&\multicolumn{3}{c}{SWE}\\
		\toprule
		&&$P/S$&$PS$&$SS$\\
		\toprule
		\input{tab_fraction_SWE}
		\bottomrule
		\end{tabularx}
\begin{flushright}
	\end{flushright}
	\end{minipage}\hspace{-0.13cm}
\begin{minipage}{.26\linewidth}	\vspace*{-.25cm}
		\vspace{-0.13cm}\begin{tabularx}{1\linewidth}{cccc}
		&\multicolumn{3}{c}{DEN}\\
		\toprule
		&$P/S$&$PS$&$SS$\\
		\toprule
		\input{tab_fractions_DEN}
		\bottomrule
		\end{tabularx}
	\end{minipage}
		\begin{minipage}{.99\linewidth}
			\vspace*{-.32cm}
	\footnotesize{Notes: $P:$ PES vacancies, $S:$ survey vacancies, $PS:$ share of all PES vacancies, and $SS:$ share of all survey vacancies. For Sweden the quartiles refer to value-added per worker. For Denmark the quartiles refer to the gross salary per worker. }	
	\end{minipage}
\end{center}
\end{table}

Table \ref{tab:fracshares} does not highlight that a given aggregate fraction of PES to survey vacancies does not need to stem from the same fraction in all firms. Thus, in Table \ref{tab:fracsonline} we look more closely at the relationship between survey and PES vacancies at the firm level, as first presented in Figure \ref{fig:scatterplot}. In columns (1) and (5) we report the number of observations where the two measures are identical. Columns (2) and (6) are similar but involve a restriction the sample to observations where at least one of two measure is  non-zero. In columns (3) and (7) we report the fraction of observations with a higher number of PES than survey vacancies. Columns (4) and (8) report the fraction of all PES vacancies that do not have corresponding number of survey vacancy registered. The fractions are computed in relation to all survey vacancies: \emph{E.g.} if a firm has posted 10 PES and 2 survey vacancies in one period, then this firm contribute with 8 to the numerator of the NC measure in columns (4) and (8). 

% TABLE 6  Fractions on line
\begin{sidewaystable}
\begin{center}
			\caption{Fractions on line}
			\label{tab:fracsonline}
		\begin{minipage}{.60\linewidth}
\begin{flushright}
		\begin{tabularx}{1\linewidth}{cXcccc}
		&&\multicolumn{4}{c}{SWE}\\
		\toprule
		&&(1)&(2)&(3)&(4)\\
		&&$P=S$&\normalsize{$P=S$}&$P>S$&$NC$\\
		\toprule
		\input{tab_fraction_online_SWE}
		\midrule
		&Vacancy sample&All&$S>0$&All&All\\
		\bottomrule
		\end{tabularx}
	\end{flushright}
	\end{minipage}\hspace{-0.12cm}
		\begin{minipage}{.275\linewidth}
		\begin{tabularx}{1\linewidth}{ccccc}
		\multicolumn{5}{c}{DEN}\\
		\toprule
		&(5)&(6)&(7)&(8)\\
		&$P=S$&$P=S$&$P>S$&$NC$\\
		\toprule
		\input{tab_online_DEN}
		\midrule
		&All&$S>0$&All&All\\
		\bottomrule
		\end{tabularx}
	\end{minipage}
		\begin{minipage}{.875\linewidth}
	\footnotesize{Notes: $P:$ PES vacancies, $S:$ survey vacancies and $NC:$ Fraction of PES vacancies with no survey vacancy. 
(1) and (5) report the number of observations where the two measures are identical, columns (2) and (6) do the same but with the sampled restricted to observations where at least one of two measure is non-zero. Columns (3) and (7) report the fraction of observations with a higher number of PES than survey vacancies. Columns (4) and (8) reports the fraction of all PES vacancies, which do not have a number corresponding survey vacancy registered as a share of all survey vacancies.}	
	\end{minipage}
\end{center}
\end{sidewaystable}

In both countries the share of observations with an equal number of PES and survey vacancies is around 80\%. However, a large share of these observations are located at the origin. Indeed, only 7\% and 3\% of all non-zero observations are identical. As noted in the discussion of Figure \ref{fig:scatterplot} above, there are also observations with a higher number of PES than survey vacancies. These account for 4\% of all observations in Sweden and 2\% in Denmark. 

The findings so far have shown that the aggregate number of PES vacancies constitutes a fraction of the survey vacancies. This may lead one to presume that PES vacancies simply form a subset of survey vacancies. In column (4) and (8) we show that this presumption is false. Indeed, 61\% and 64\% of the vacancies in the PES are not present in the survey. 

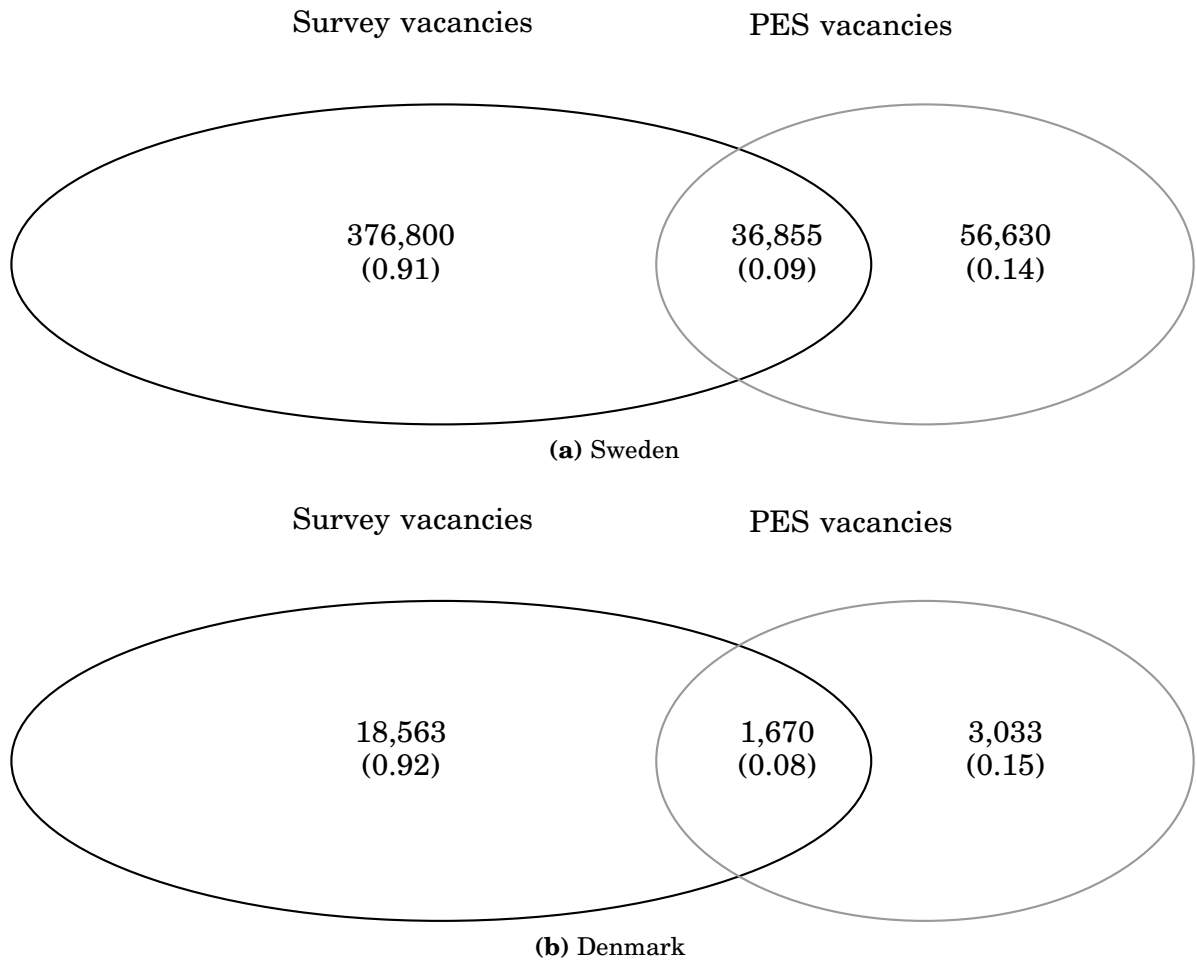
\begin{figure}
	\subfloat[Sweden]{\resizebox{1\linewidth}{!}{
\begin{tikzpicture}
  \node[text width=6cm] at (30pt,90pt)  { Survey vacancies};
  \draw[black,thick] (0,0) ellipse (160pt and 60pt);
  \node[text width=6cm] at (200pt,90pt)  {  PES vacancies};
  \draw[gray!80!white,thick,label=above:PES] (180pt,0) ellipse (100pt and 60pt);
  \node[text width=3cm, text centered] at (-15pt,10pt)  {  376,800};
  \node[text width=3cm, text centered] at (-15pt,-2pt)  {  (0.91)};
  \node[text width=3cm, text centered] at (125pt,10pt)  {  36,855};
  \node[text width=3cm, text centered] at (125pt,-2pt)  {  (0.09)};
  \node[text width=3cm, text centered] at (210pt,10pt)  {  56,630};
  \node[text width=3cm, text centered] at (210pt,-2pt)  { (0.14)};
\end{tikzpicture}\\
}
}\\
	\subfloat[Denmark]{\resizebox{1\linewidth}{!}{
\begin{tikzpicture}
  \node[text width=6cm] at (30pt,90pt)  { Survey vacancies};
  \draw[black,thick] (0,0) ellipse (160pt and 60pt);
  \node[text width=6cm] at (200pt,90pt)  {  PES vacancies};
  \draw[gray!80!white,thick,label=above:PES] (180pt,0) ellipse (100pt and 60pt);
  \node[text width=3cm, text centered] at (-15pt,10pt)  {  18,563};
  \node[text width=3cm, text centered] at (-15pt,-2pt)  {  (0.92)};
  \node[text width=3cm, text centered] at (125pt,10pt)  {  1,670};
  \node[text width=3cm, text centered] at (125pt,-2pt)  {  (0.08)};
  \node[text width=3cm, text centered] at (210pt,10pt)  {  3,033};
  \node[text width=3cm, text centered] at (210pt,-2pt)  { (0.15)};
\end{tikzpicture}\\
}}
\caption{Vacancies in the Survey, the PES and total vacancies. The uppermost numbers are the number of vacancies. The numbers in parenthesis indicate the fraction relative to the total number of survey vacancies.}
\label{fig:venn_diagram}
\end{figure}

Figure \ref{fig:venn_diagram} shows the overlap and dis joint sub-sets of the PES and survey vacancies. Slightly more than 90\% of the survey vacancies are not present in the PES database and the remaining 10\% are present in both databases. Finally, the number of survey vacancies would grow by approximately 15\% if all PES vacancies were also present in the survey. That is, if the survey truly was a super-set of the PES vacancies. 

To assess whether the ratio of total vacancies to survey vacancies varies systematically across firms  we regress the ratio on observable firm characteristics in  Table \ref{tab:regs}. Columns (2) and (4) show that for both countries the ratio increases in firm size. Also for both countries firms within \emph{Public and personal services}  have significantly higher ratios than for the reference group, \emph{Energy and water supply}.

\section{Aggregate implications}
\markboth{\MakeUppercase{Two Vacancy Measures}}{\MakeUppercase{Aggregate implications}}
\label{sec:agg_implications}

Above we have shown that a substantial amount of vacancies are present in the PES but not in the survey data. This means, that one cannot simply interpret the latter as a super-set of the former. This has implications on the aggregate level, as one has to take into account the vacancies present in the PES, but not in the survey, in order to get closer to the true number of vacancies in the economy. 

\begin{figure}[!ht]
	\subfloat[Sweden]{\includegraphics[width=0.5\linewidth]{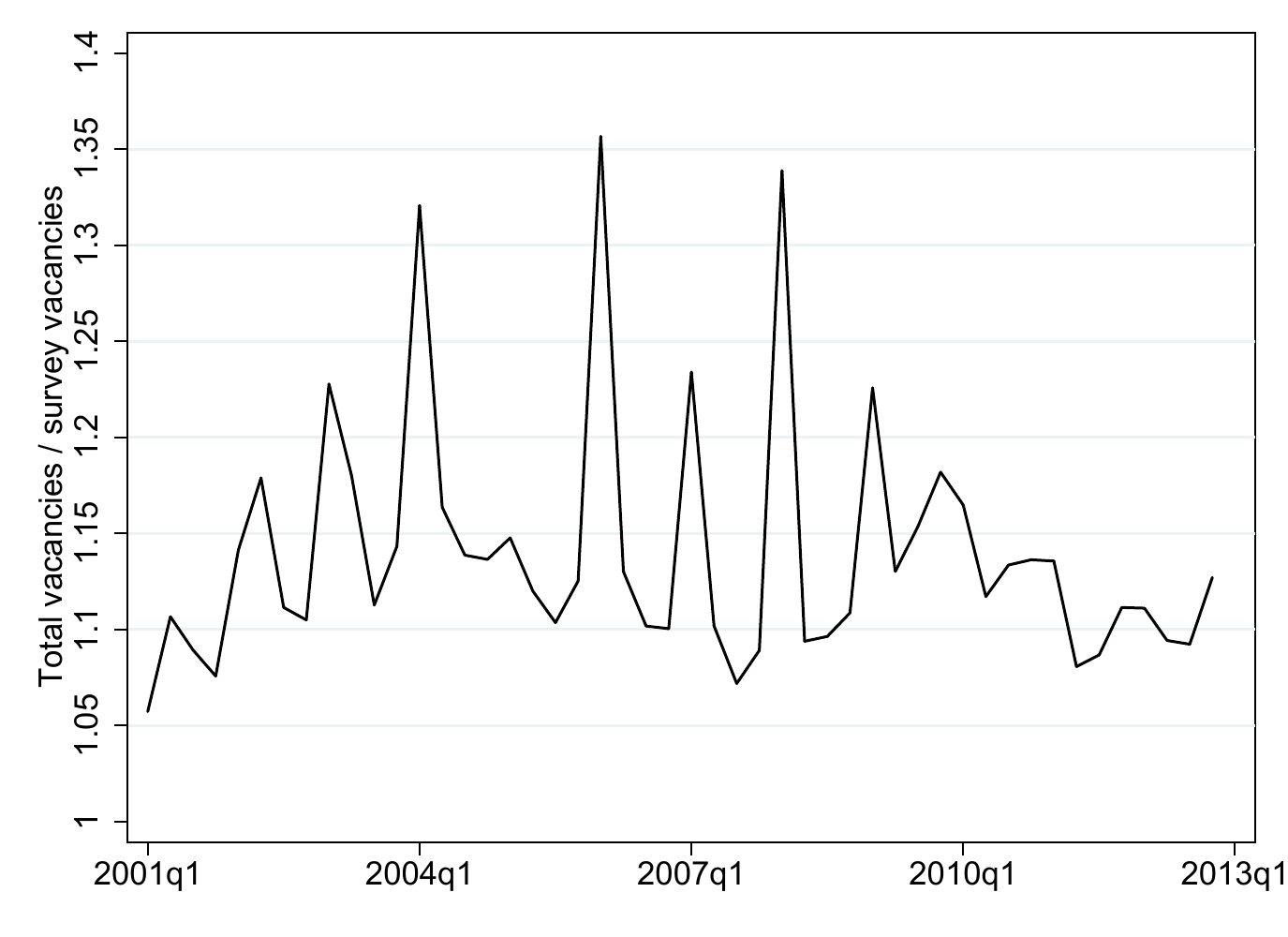}}
	\subfloat[Denmark]{\includegraphics[width=0.5\linewidth]{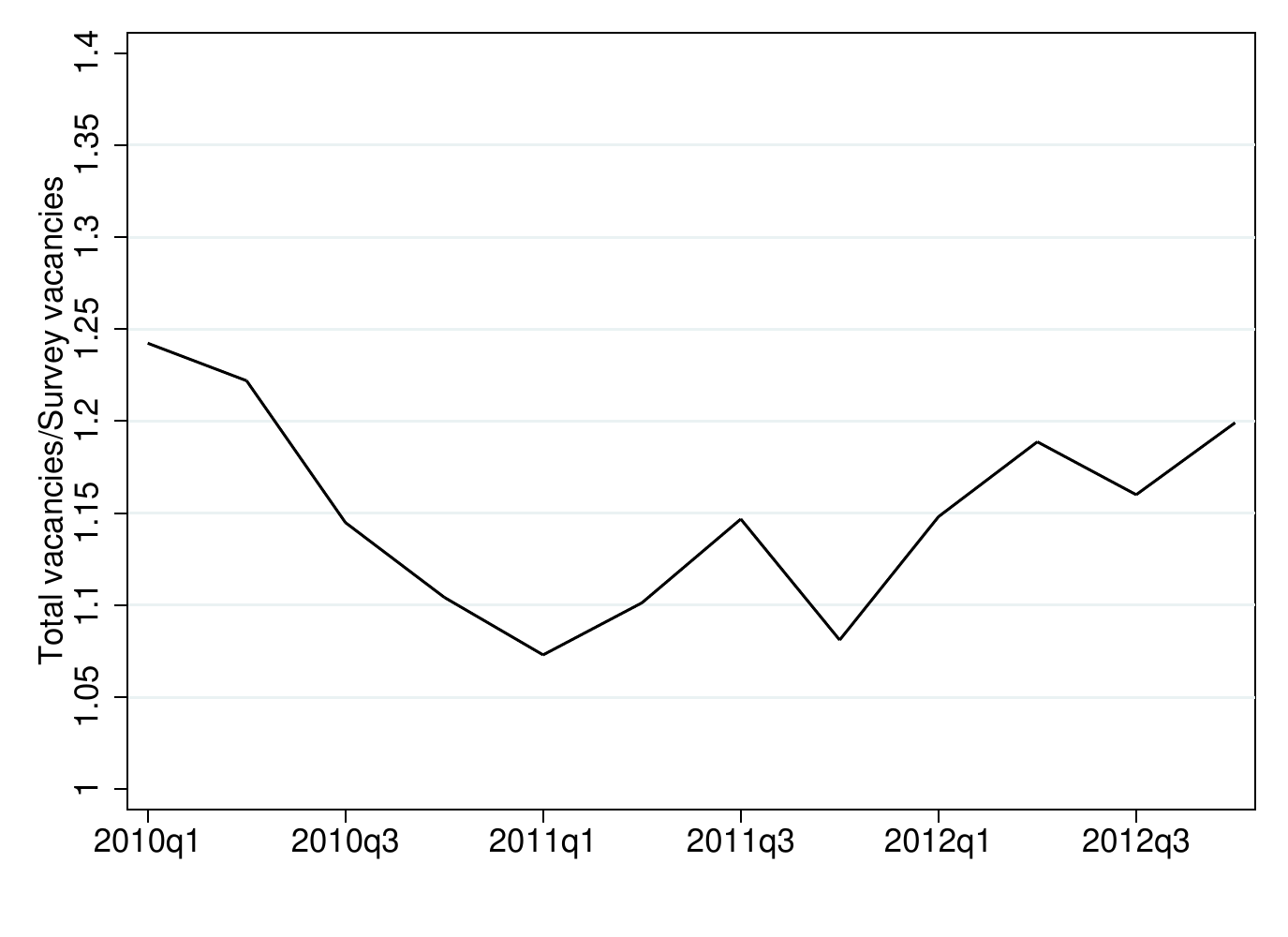}}\\
	\caption{Aggregated total vacancies related to aggregated survey vacancies}
	\label{fig:aggregatedratio}
\end{figure}

In Figure \ref{fig:aggregatedratio} we report the full set of unique vacancies in our database to the number of vacancies counted in the survey. The full set of unique vacancies is computed by adding the vacancies unique to the PES to the total number of survey vacancies. For both countries, this ratio varies in the range 1 to 1.3. In Appendix Figure \ref{fig:aggregated} we report the underlying times series.

To assess whether the ratio varies systematically over the business cycle,  Table \ref{tab:regs}  reports the coefficients from a regression of the ratio on the rate of unemployment. For Sweden the ratio varies negatively with the rate of unemployment, while this relationship is insignificant in Denmark. The latter is likely due to a shorter time period. In other words, when labor market conditions improve the level of mis-measurement tends to increase.

\begin{table}[!ht]
\begin{center}
			\caption{Regressions results - Dependent variable: The ratio of total vacancies to survey vacancies}
			\label{tab:regs}
			% \\
		\resizebox{1\linewidth}{!}{\begin{minipage}{.8\linewidth}
\begin{flushright}
	\begin{tabularx}{1\linewidth}{Xccc}
	\toprule
		&\multicolumn{3}{c}{SWE}\\
		&(1)&(2)&(3)\\
		\toprule
		\input{tab_ratioreg_SWE}
		\bottomrule
		\end{tabularx}
	\end{flushright}
	\end{minipage}\hspace{-0.12cm}
		\begin{minipage}{.35\linewidth}
		\begin{tabularx}{1\linewidth}{cccc}
		\toprule
		&\multicolumn{3}{c}{DEN}\\
		&(4)&(5)&(6)\\
		\toprule
		\input{tab_ratioreg_DEN}
		\bottomrule
		\end{tabularx}
	\end{minipage}}
		\begin{minipage}{1\linewidth}
	\footnotesize{Notes:  The columns show the point-estimates for $\alpha$ from estimating the following model: $ratio_{it}=\alpha_0+\alpha_1 une_{it}+\bm{\beta}\bm{X}_{it}+ \varepsilon_{it}$ using ordinary least squares. Where $ratio$ is the ratio of  total vacancies to survey vacancies and $une$ is the unemployment rate and $\bm{X}$ is the set of indicators capturing firm-level characteristics.  The reference group for industry is Energy and water supply. Standard errors clustered on the firm-level in parenthesis. Confidence levels are indicated as follows: $^{*}$ $p<0.1$, $^{**}$ $p<0.05$, and $^{***}$ $p<0.01$.}	
	\end{minipage}
\end{center}
\end{table}

\FloatBarrier
\section{Concluding remarks}
\markboth{\MakeUppercase{Two Vacancy Measures}}{\MakeUppercase{Concluding remarks}}
\label{sec:paper3:conclusion}

Using linked survey- and register-data for both Sweden and Denmark we have investigated the firm level relationship between these two commonly used vacancy measures. One is from a survey compiled by national statistical offices, while the other is register data from public employment services. 

It it well known that the survey based measure covers a larger set of vacancies. Consistently with this we find that the sum of PES vacancies only constitutes around a quarter of all survey vacancies in our sample. However, the firm level structure of our data allows us to establish that the survey vacancies are not a super-set of vacancies in the PES. In particular, we find that the survey based measure must be adjusted by factor of approximately 1.2 to represent the full set of unique vacancies in these two measures. 

This mis-measurement of vacancies is not necessarily a problem if is unrelated to underlying firm characteristics and business cycle fluctuations. Shifts in the number of survey vacancies would still be informative in this case about changes in the labor market conditions. However, we show that the mis-measurement varies systematically over time and across firm characteristics. This implies that variation in survey vacancies is not only caused by changes in the actual number of vacancies, but also by changes in the level of mis-measurement.

We find that the ratio of total vacancies to survey vacancies varies from less than 1.1 to more than 1.25 over time. In other words, an observed increase in survey vacancies of 15\% could solely be the outcome of changes in mis-measurement. The variation in mis-measurement over time might be a result of changes in the underlying composition of job vacancies across firm type, as the fraction of PES vacancies not included in the survey data varies from less than 60 to more than 80 percent across industries and firm size. 

The degree of mis-measurement has similarities across the two countries. The number of PES vacancies constitutes about one quarter of the total number of survey vacancies in both Sweden and Denmark, but about 60\% of all PES vacancies are not included in the survey data in both countries. With respect to firm level characteristics, however, the patterns varies between the two countries. 

Let us conclude by suggesting some questions for future research. First, it is important to note that our findings are specific to a certain subset of firms, and not necessarily representative for the economy as a whole. It would be important to assess whether our findings carry over to a broader set of firms in the economy. Second, it would interesting to investigate how the coefficients in the matching functions would change when using the number of unique vacancies in the survey and PES as a alternative vacancy measure.

\FloatBarrier
\pagebreak
\bibliography{lib}

\clearpage
\appendix 
\renewcommand\thefigure{\thesection\arabic{figure}}    
\renewcommand\thetable{\thesection\arabic{table}}    
\section{Appendix}
\setcounter{figure}{0}    
\setcounter{table}{0}

% Representativity
\begin{table}[ht!]
\begin{center}
			\caption{Firm distribution in economy (E) and sample (S)}
			\label{tab:representiveness}
		\begin{minipage}{.73\linewidth}
			\begin{tabularx}{1\linewidth}{cXcccc}
		&&\multicolumn{3}{c}{SWE}\\
		\toprule
		&&$E$&$S$&\\
		\toprule
		\input{tab_represent_SWE}
		\\
		\bottomrule
		\end{tabularx}
\begin{flushright}
	\end{flushright}
	\end{minipage}\hspace{-0.13cm}
\begin{minipage}{.26\linewidth}	\vspace*{-.25cm}
		\vspace{-0.13cm}\begin{tabularx}{1\linewidth}{cccc}
		&\multicolumn{3}{c}{DEN}\\
		\toprule
		&$E$&$S$&\\
		\toprule \\
		\input{tab_represent_DEN} 
		\\
		\bottomrule
		\end{tabularx}
	\end{minipage}
		\begin{minipage}{.99\linewidth}
			\vspace*{-.32cm}
	\footnotesize{Notes: $E:$ Distribution in the entire according to the Firm Database in Sweden and IDAN in Denmark, $S:$ distribution in our sample, The chi-square test is a test for the two distributions being equal. For Sweden the quartiles refer to value-added per worker. For Denmark the quartiles refer to the gross salary per worker. }	
	\end{minipage}
\end{center}
\end{table}

\begin{figure}[!ht]
	\subfloat[Sweden]{\includegraphics[width=0.5\linewidth]{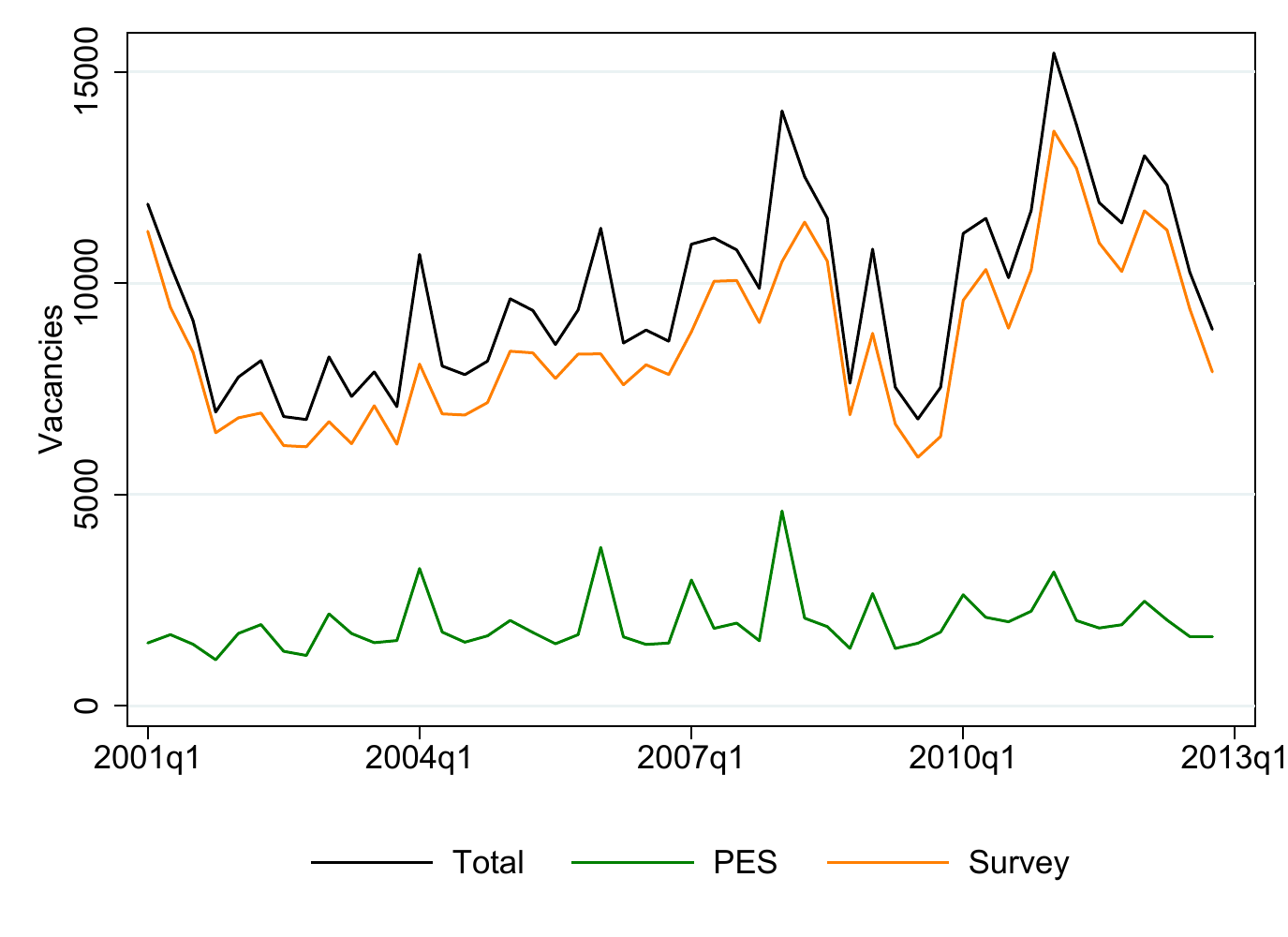}}
	\subfloat[Denmark]{\includegraphics[width=0.5\linewidth]{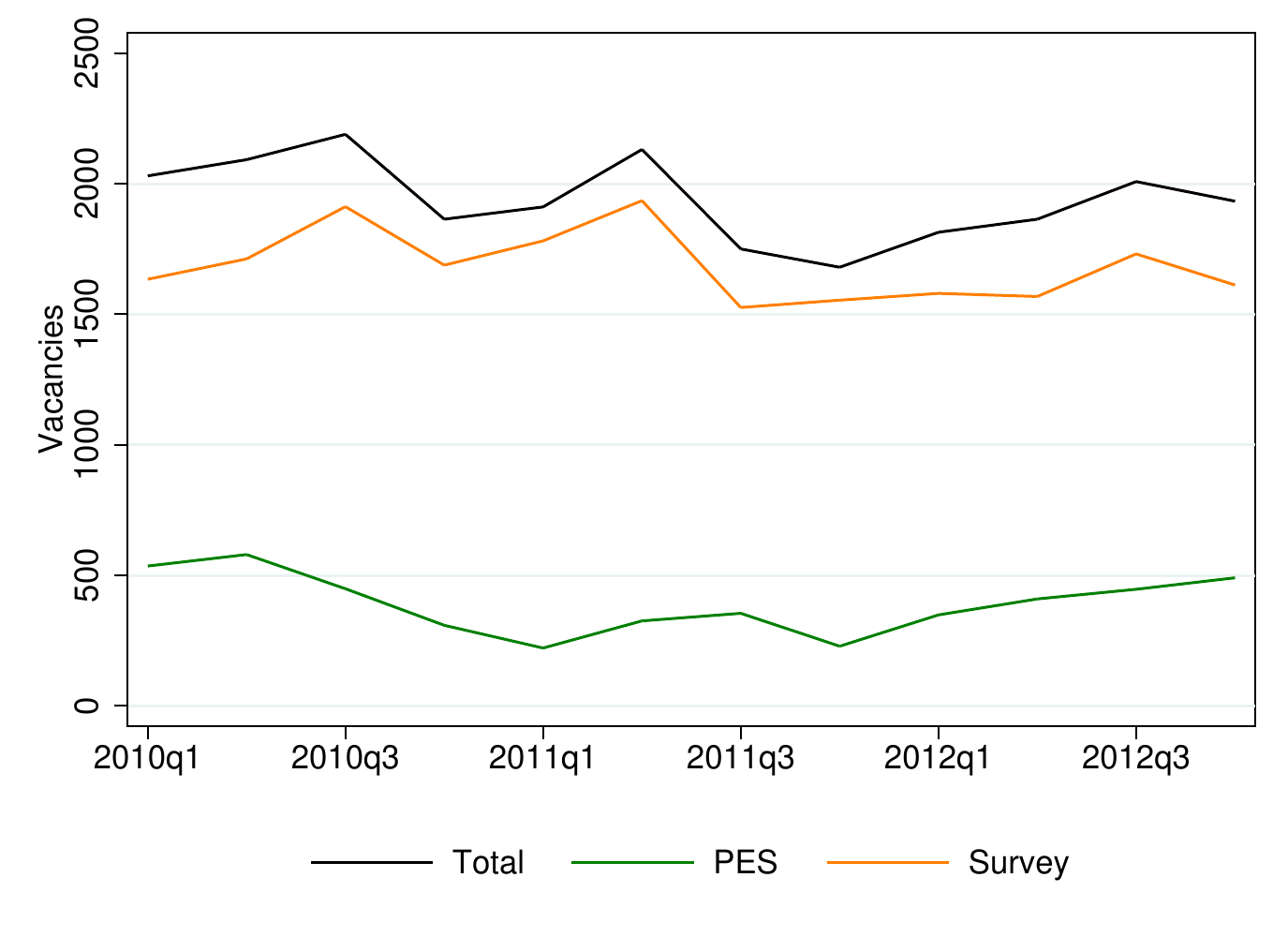}}\\
	\caption{Aggregated total vacancies, survey vacancies and PES vacancies}
	\label{fig:aggregated}
\end{figure}

\begin{table}[!ht]
\begin{center}
			\caption{Dependence on missing firm-identifiers: The ratio of total vacancies to survey vacancies}
			\label{tab:regs_before_2009}
			% \\
\begin{flushright}
	\resizebox{\linewidth}{!}{
		\begin{tabularx}{1.06\linewidth}{Xcccccc}
	\toprule
		&\multicolumn{6}{c}{SWE}\\
		&(1)&(2)&(3)&(4)&(5)&(6)\\
		\toprule
		\input{tab_ratioreg_SWE_from_2009}
		Include pre-2009?&YES&NO&YES&NO&YES&NO\\
		\bottomrule
		\end{tabularx}
	}
	\end{flushright}
	\begin{minipage}{\linewidth}
	\footnotesize{Notes:  The columns show the point-estimates for $\alpha$ from estimating the following model: $ratio_{it}=\alpha_0+\alpha_1 une_{it}+\bm{\beta}\bm{X}_{it}+ \varepsilon_{it}$ using ordinary least squares. Where $ratio$ is the ratio of  total vacancies to survey vacancies and $une$ is the unemployment rate and $\bm{X}$ is the set of indicators capturing firm-level characteristics.  The reference group for industry is Energy and water supply. Standard errors clustered on the firm-level in parenthesis. Confidence levels are indicated as follows: $^{*}$ $p<0.1$, $^{**}$ $p<0.05$, and $^{***}$ $p<0.01$.}
		\end{minipage}	
\end{center}
\end{table}

\end{document}

%% file: tab_selection_SWE.tex
Full sample &852,156& 1.00 &    2.0\\
Information on \# of plants &726,058&0.85&    2.1\\
One plant per firm &487,667&0.57&    0.8\\
Covariates available &366,610&0.43&    0.9\\

%% file: tab_selection_DEN.tex
63,722&1.00&1.2\\
61,496&0.97&1.3\\
41,731&0.65&0.5\\
41,668&0.65&0.5\\

%% file: tab_descriptives_SWE.tex
&Farming, fishery and mining &0.06&0.24\\
&Manufactoring &0.16&0.37\\
&Energy and water supply &0.01&0.09\\
&Construction &0.09&0.29\\
&Trade, Hotel and Restaurants &0.20&0.40\\
&Transportation, mail and telecom &0.05&0.22\\
&Finance and business services &0.25&0.43\\
&Public services and  personal services &0.17&0.38\\
\multicolumn{4}{l}{\emph{Firm size}}\\
&No. of employees &16.6&43.7\\
&0-5 &0.30&0.46\\
&6-10 &0.11&0.31\\
&11-50 &0.16&0.36\\
&50+ &0.43&0.50\\
\multicolumn{4}{l}{\emph{Value-added/salary}}\\
&Value-added per worker (1,000 EUR) &  64.1& 134.0\\
&Gross salary per worker (1,000 EUR) \\

%% file: tab_descriptives_DEN.tex
\\
0.03&0.18\\
0.15&0.36\\
0.15&0.35\\
0.31&0.46\\
0.11&0.31\\
0.05&0.23\\
0.14&0.35\\
0.06&0.24\\
\\
26.2&56.4\\
0.42&0.49\\
0.14&0.35\\
0.29&0.45\\
0.14&0.35\\
\\
\\
  44.4&  51.3\\

%% file: tab_vac_survey_PES_cross_SWE.tex
&0 & 0.771& 0.063& 0.032& 0.015& 0.033&\\
&1 & 0.026& 0.011& 0.008& 0.005& 0.014&\\
&2 & 0.004& 0.001& 0.001& 0.001& 0.003&\\
&3 & 0.002& 0.000& 0.000& 0.000& 0.001&\\
&4+ & 0.004& 0.001& 0.001& 0.000& 0.002&\\

%% file: tab_vac_PES_SURV_DEN.tex
 &0 &0.832&0.069&0.031&0.013&0.025&\\
 &1 &0.010&0.004&0.003&0.001&0.002&\\
 &2 &0.002&0.001&0.001&0.000&0.001&\\
 &3 &0.001&0.000&0.000&0.000&0.001&\\
 &4+ &0.000&0.000&0.000&0.000&0.000&\\

%% file: tab_fraction_SWE.tex
& All                                    & 0.23  & 1.00 & 1.00\\
\multicolumn{4}{l}{\emph{By industry}} \\
& Farming, fishery and mining            & 0.63  & 0.02 & 0.01\\
& Manufacturing                          & 0.20  & 0.30 & 0.33\\
& Energy and water supply                & 0.10  & 0.01 & 0.02\\
& Construction                           & 0.21  & 0.02 & 0.02\\
& Trade, Hotel and Restaurants           & 0.25  & 0.11 & 0.10\\
& Transportation, mail and telecom       & 0.21  & 0.06 & 0.06\\
& Finance and business services          & 0.23  & 0.23 & 0.23\\
& Public services and  personal services & 0.24  & 0.25 & 0.23\\
\multicolumn{4}{l}{\emph{By firm size}} \\
& 0-5                                    & 0.38  & 0.02 & 0.01\\
& 6-10                                   & 0.38  & 0.03 & 0.02\\
& 11-50                                  & 0.37  & 0.13 & 0.08\\
& 50+                                    & 0.21  & 0.82 & 0.89\\
\multicolumn{4}{l}{\emph{By value-added/salary per worker}} \\
& Quartile 1                             & 0.35  & 0.23 & 0.15\\
& Quartile 2                             & 0.36  & 0.20 & 0.13\\
& Quartile 3                             & 0.20  & 0.17 & 0.19\\
& Quartile 4                             & 0.17  & 0.22 & 0.29\\

%% file: tab_fractions_DEN.tex
&                  0.23&         1.00&1.00\\
\\
&                  0.09&      0.01&0.03\\
&                  0.10&         0.11&0.25\\
&                  0.09&         0.02&0.06\\
&                  0.08&         0.06&0.17\\
&                  0.36&         0.10&0.07\\
&                  0.06&         0.03&0.12\\
&                  0.33&         0.19&0.13\\
&                  0.74&         0.46&0.14\\
\\
&                  0.13&      0.03&0.05\\
&                  0.21&         0.03&0.03\\
&                  0.45&         0.34&0.18\\
&                  0.19&         0.59&0.74\\
\\
&                  0.82&      0.50&0.14\\
&                  0.46&         0.24&0.12\\
&                  0.13&         0.12&0.22\\
&                  0.05&         0.10&0.49\\

%% file: tab_fraction_online_SWE.tex
&All &0.78&0.07&0.04&0.61\\
\multicolumn{6}{l}{\emph{By industry}} \\
&Farming, fishery and mining &0.93&0.06&0.01&0.83\\
&Manufactoring &0.72&0.07&0.04&0.57\\
&Energy and water supply &0.75&0.07&0.04&0.49\\
&Construction &0.90&0.04&0.01&0.52\\
&Trade, Hotel and Restaurants &0.82&0.08&0.04&0.65\\
&Transportation, mail and telecom &0.77&0.04&0.04&0.65\\
&Finance and business services &0.79&0.05&0.04&0.63\\
&Public services and  personal services &0.76&0.10&0.06&0.59\\
\multicolumn{6}{l}{\emph{By firmsize}} \\
&0-5 &0.97&0.06&0.01&0.82\\
&6-10 &0.91&0.08&0.02&0.74\\
&11-50 &0.83&0.08&0.04&0.73\\
&50+ &0.61&0.06&0.07&0.58\\
\multicolumn{6}{l}{\emph{By value-added per worker quartile}} \\
&Quartile 1 &0.83&0.08&0.06&0.64\\
&Quartile 2 &0.81&0.09&0.05&0.65\\
&Quartile 3 &0.77&0.07&0.04&0.57\\
&Quartile 4 &0.72&0.05&0.03&0.61\\

%% file: tab_online_DEN.tex
&                  0.84&     0.03&     0.02&     0.64\\
\\
&                  0.84&     0.03&     0.02&     0.76\\
&                  0.76&     0.03&     0.01&     0.40\\
&                  0.91&     0.02&     0.01&     0.61\\
&                  0.87&     0.02&     0.01&     0.59\\
&                  0.84&     0.04&     0.03&     0.82\\
&                  0.68&     0.01&     0.01&     0.54\\
&                  0.81&     0.03&     0.01&     0.86\\
&                  0.77&     0.04&     0.09&     0.60\\
\\
&                  0.96&     0.01&     0.00&     0.67\\
&                  0.92&     0.02&     0.01&     0.79\\
&                  0.84&     0.03&     0.02&     0.82\\
&                  0.56&     0.03&     0.04&     0.54\\
\\
&                  0.91&     0.04&     0.03&     0.68\\
&                  0.90&     0.05&     0.02&     0.71\\
&                  0.81&     0.03&     0.01&     0.46\\
&                  0.71&     0.01&     0.01&     0.52\\

%% file: tab_ratioreg_SWE.tex
Unemployment        &       -0.14         &                     &       -0.82\sym{**} \\
                    &      (0.18)         &                     &      (0.36)         \\
Farming, fishery and mining&                     &        0.02\sym{**} &        0.02\sym{**} \\
                    &                     &      (0.01)         &      (0.01)         \\
Manufactoring       &                     &        0.02         &        0.02         \\
                    &                     &      (0.02)         &      (0.02)         \\
Construction        &                     &        0.01         &        0.02         \\
                    &                     &      (0.01)         &      (0.01)         \\
Trade, Hotel and Restaurants&                     &        0.02\sym{*}  &        0.02\sym{*}  \\
                    &                     &      (0.01)         &      (0.01)         \\
Transportation, mail and telecom&                     &        0.01\sym{*}  &        0.02\sym{*}  \\
                    &                     &      (0.01)         &      (0.01)         \\
Finance and business services&                     &        0.02\sym{***}&        0.02\sym{***}\\
                    &                     &      (0.01)         &      (0.01)         \\
Public services and  personal services&                     &        0.04\sym{***}&        0.04\sym{***}\\
                    &                     &      (0.01)         &      (0.01)         \\
6-10 employees      &                     &        0.00\sym{*}  &        0.00\sym{**} \\
                    &                     &      (0.00)         &      (0.00)         \\
11-50 employees     &                     &        0.00\sym{***}&        0.01\sym{***}\\
                    &                     &      (0.00)         &      (0.00)         \\
50+ employees       &                     &        0.06\sym{***}&        0.06\sym{***}\\
                    &                     &      (0.01)         &      (0.01)         \\
VA/Salary per worker, quartile 2&                     &       -0.00         &       -0.00         \\
                    &                     &      (0.01)         &      (0.01)         \\
VA/Salary per worker, quartile 3&                     &       -0.01\sym{**} &       -0.01\sym{**} \\
                    &                     &      (0.00)         &      (0.00)         \\
VA/Salary per worker, quartile 4&                     &        0.01         &        0.01         \\
                    &                     &      (0.01)         &      (0.01)         \\
Constant            &        1.03\sym{***}&        0.98\sym{***}&        1.04\sym{***}\\
                    &      (0.01)         &      (0.01)         &      (0.02)         \\\midrule
Observations        &      470,073         &      352,126         &      352,126         \\

%% file: tab_ratioreg_DEN.tex
                    &       -0.01         &                     &        0.02         \\
                    &      (0.48)         &                     &      (0.50)         \\
                    &                     &       -0.01         &       -0.01         \\
                    &                     &      (0.01)         &      (0.01)         \\
                    &                     &       -0.00         &       -0.00         \\
                    &                     &      (0.00)         &      (0.00)         \\
                    &                     &       -0.00         &       -0.00         \\
                    &                     &      (0.00)         &      (0.00)         \\
                    &                     &        0.01         &        0.01         \\
                    &                     &      (0.02)         &      (0.02)         \\
                    &                     &        0.01\sym{*}  &        0.01\sym{*}  \\
                    &                     &      (0.00)         &      (0.00)         \\
                    &                     &        0.03         &        0.03         \\
                    &                     &      (0.02)         &      (0.02)         \\
                    &                     &        0.12\sym{***}&        0.12\sym{***}\\
                    &                     &      (0.04)         &      (0.04)         \\
                    &                     &        0.00         &        0.00         \\
                    &                     &      (0.00)         &      (0.00)         \\
                    &                     &        0.03\sym{*}  &        0.03\sym{*}  \\
                    &                     &      (0.01)         &      (0.01)         \\
                    &                     &        0.04\sym{***}&        0.04\sym{***}\\
                    &                     &      (0.01)         &      (0.01)         \\
                    &                     &       -0.00         &       -0.00         \\
                    &                     &      (0.02)         &      (0.02)         \\
                    &                     &       -0.02         &       -0.02         \\
                    &                     &      (0.02)         &      (0.02)         \\
                    &                     &       -0.02         &       -0.02         \\
                    &                     &      (0.02)         &      (0.02)         \\

		&        1.01\sym{***}&        1.00\sym{***}&        0.99\sym{***}\\
                    &      (0.04)         &      (0.01)         &      (0.04)         \\\midrule
&       41,152         &       38,974         &       38,974         \\

%% file: tab_represent_SWE.tex
\\
\multicolumn{4}{l}{\emph{By industry}} \\
&Farming, fishery and mining &0.22&0.06\\
&Manufacturing &   0.07&     0.20\\
&Energy and water supply &0.00&0.01\\
&Construction &    0.08&     0.08\\
&Trade, Hotel and Restaurants &0.17&0.19\\
&Transportation, mail and telecom &0.04&0.05\\
&Finance and business servce &0.27&0.24\\
&Public and  personal services &0.15&0.17\\
&$\chi^2$ test&&&                      0.00\\
\multicolumn{4}{l}{\emph{By firm size}} \\
&-5 &              0.57&     0.36\\
&5-10 &            0.18&     0.17\\
&10-50 &           0.20&     0.31\\
&50 &              0.05&     0.17\\
&$\chi^2$ test&&&                      0.00\\
\multicolumn{4}{l}{\emph{By value-added/salary per worker}} \\
&Quartile 1 &      0.26&     0.15\\
&Quartile 2 &      0.25&     0.23\\
&Quartile 3 &      0.25&     0.29\\
&Quartile 4 &      0.24&     0.32\\
&$\chi^2$ test&&&                      0.00\\

%% file: tab_represent_DEN.tex
\\
&                  0.06&     0.03\\
&                  0.05&     0.15\\
&                  0.09&     0.14\\
&                  0.20&     0.29\\
&                  0.07&     0.10\\
&                  0.04&     0.05\\
&                  0.11&     0.13\\
&                  0.18&     0.05\\
&&&                                    0.00\\
\\
&                  0.72&     0.39\\
&                  0.12&     0.14\\
&                  0.13&     0.30\\
&                  0.02&     0.17\\
&&&                                    0.00\\
\\
&                  0.25&     0.14\\
&                  0.25&     0.20\\
&                  0.25&     0.29\\
&                  0.25&     0.37\\
&&&                                    0.00\\ 

%% file: tab_ratioreg_SWE_from_2009.tex
Unemployment        &                     &                     &       -0.14         &       -0.60         &                     &                     \\
                    &                     &                     &      (0.18)         &      (0.49)         &                     &                     \\
Farming, fishery and mining&                     &                     &                     &                     &        0.02\sym{**} &        0.02\sym{***}\\
                    &                     &                     &                     &                     &      (0.01)         &      (0.00)         \\
Manufactoring       &                     &                     &                     &                     &        0.02         &        0.01         \\
                    &                     &                     &                     &                     &      (0.02)         &      (0.01)         \\
Construction        &                     &                     &                     &                     &        0.01         &        0.01\sym{*}  \\
                    &                     &                     &                     &                     &      (0.01)         &      (0.00)         \\
Trade, Hotel and Restaurants&                     &                     &                     &                     &        0.02\sym{*}  &        0.01\sym{*}  \\
                    &                     &                     &                     &                     &      (0.01)         &      (0.00)         \\
Transportation, mail and telecom&                     &                     &                     &                     &        0.01\sym{*}  &        0.01         \\
                    &                     &                     &                     &                     &      (0.01)         &      (0.01)         \\
Finance and business services&                     &                     &                     &                     &        0.02\sym{***}&        0.03\sym{***}\\
                    &                     &                     &                     &                     &      (0.01)         &      (0.01)         \\
Public services and  personal services&                     &                     &                     &                     &        0.04\sym{***}&        0.04\sym{***}\\
                    &                     &                     &                     &                     &      (0.01)         &      (0.01)         \\
6-10 employees      &                     &                     &                     &                     &        0.00\sym{*}  &        0.00\sym{***}\\
                    &                     &                     &                     &                     &      (0.00)         &      (0.00)         \\
11-50 employees     &                     &                     &                     &                     &        0.00\sym{***}&        0.01\sym{***}\\
                    &                     &                     &                     &                     &      (0.00)         &      (0.00)         \\
50+ employees       &                     &                     &                     &                     &        0.06\sym{***}&        0.04\sym{***}\\
                    &                     &                     &                     &                     &      (0.01)         &      (0.01)         \\
VA/Salary per worker, quartile 2&                     &                     &                     &                     &       -0.00         &       -0.01\sym{**} \\
                    &                     &                     &                     &                     &      (0.01)         &      (0.00)         \\
VA/Salary per worker, quartile 3&                     &                     &                     &                     &       -0.01\sym{**} &       -0.01\sym{**} \\
                    &                     &                     &                     &                     &      (0.00)         &      (0.01)         \\
VA/Salary per worker, quartile 4&                     &                     &                     &                     &        0.01         &       -0.00         \\
                    &                     &                     &                     &                     &      (0.01)         &      (0.01)         \\
Constant            &        1.02\sym{***}&        1.02\sym{***}&        1.03\sym{***}&        1.07\sym{***}&        0.98\sym{***}&        0.99\sym{***}\\
                    &      (0.00)         &      (0.00)         &      (0.01)         &      (0.04)         &      (0.01)         &      (0.01)         \\
Observations        &      470073         &      124506         &      470073         &      124506         &      352126         &      104508         \\